# A Guide to Design Disturbance Observer

**Emre SARIYILDIZ[1]**

Affiliation: Department of System Design Engineering, Keio University, Yokohama, Japan
Address: 3-14-1 Hiyoshi, Kohoku-ku, Yokohama, Kanagawa, Japan, 223-8522, Ohnishi Laboratory, Department of System Design Engineering, Keio University
e-mail: emre@sum.sd.keio.ac.jp

**Kouhei OHNISHI[2]**

Affiliation: Department of System Design Engineering, Keio University, Yokohama, Japan
Address: 3-14-1 Hiyoshi, Kohoku-ku, Yokohama, Kanagawa, Japan, 223-8522, Ohnishi Laboratory, Department of System Design Engineering, Keio University
e-mail: ohnishi@sd.keio.ac.jp

**ABSTRACT**

*The goal of this paper is to clarify the robustness and performance constraints in the design of control systems based on disturbance observer (DOB). Although the bandwidth constraints of a DOB have long been very well-known by experiences and observations, they have not been formulated and clearly reported yet. In this regard, the Bode and Poisson integral formulas are utilized in the robustness analysis so that the bandwidth constraints of a DOB are derived analytically. In this paper, it is shown that the bandwidth of a DOB has upper and lower bounds to obtain a good robustness if the plant has non-minimum phase zero(s) and pole(s), respectively. Besides that the performance of a system can be improved by using a higher order disturbance observer (HODOB); however, the robustness may deteriorate, and the bandwidth constraints become more severe. New analysis and design methods, which provide good robustness and predefined performance criteria, are proposed for the DOB based robust control systems. The validity of the proposals are verified by simulation results.*

## 1. INTRODUCTION

---

[1] E. Sariyildiz is with the Department of System Design Engineering, Keio University, Yokohama, 223-8522, Japan (e-mail: emre@sum.sd.keio.ac.jp)
[2] Corresponding author information can be added as a footnote.

A DOB, which was proposed by K. Ohnishi et al., is a robust control tool that is used to estimate external disturbances and system uncertainties [1-3]. The estimated disturbances, which include system uncertainties, are fed-back by using an inner feed-back loop so that the robustness of a system is achieved by using a DOB [2]. Performance goals of a system are achieved by using an outer feed-back loop controller that is designed independently by considering only the nominal plant model, since a DOB can nominalize the inner-loop [4, 5]. This control structure is called as two-degrees-of-freedom control in the literature [5]. Although a DOB has been widely used in several motion control applications, e.g., robotics, industrial automation and automotive, in the last two decades, it has no systematic analysis and design methods [6-8]. Therefore, the performance and robustness of a DOB based control system highly depend on designers' own experiences.

A low pass filter (LPF) and the inverse of a nominal plant model are required to design a DOB. Although the LPF of a DOB is essential to satisfy causality in the inner-loop, it is one of the main robustness and performance limitation sources in the control systems based on DOB [9,10]. Besides that the inverse of a nominal plant model causes internal stability problem if the plant has non-minimum phase zero(s); therefore, a special consideration is required when a DOB is implemented to a non-minimum phase plant [11-13].

It is a well-known fact that a DOB can estimate disturbances precisely if they stay within the bandwidth of the DOB's LPF [14,15]. Therefore, its bandwidth is desired to set as high as possible to estimate disturbances in a wide frequency range, i.e., to improve the robustness and performance [2]. However, the bandwidth of a DOB is limited by the robustness of a system and noise, so it cannot be shaped freely [2, 10]. The noise limitation is directly related to sampling rate and measurement plants and methodology; it puts an upper bound on the bandwidth of a DOB [2]. Several researches have been reported to increase the bandwidth of a DOB by suppressing the noise of measurement [16-18]. The robustness of a DOB based control system is directly related to the dynamic characteristics of the DOB's LPF and nominal plant. They also limit the bandwidth of a DOB; however, the relation between the robustness of a system and the dynamic characteristics of the DOB's LPF and nominal plant has not been clearly reported yet [10,19,20]. Recently, it was shown by the authors that if a minimum-phase system has only real parametric uncertainties, then a DOB can guarantee the robustness of the system by increasing its bandwidth, and the stability margin of the system improves as the bandwidth of the DOB is increased [21]. However, it considers only the minimum-phase systems which have real parametric uncertainties when a first order DOB is used.

The main aim of this paper is to clarify the robustness constraints of DOB for a broad range of application area. The Bode integral formula is utilized so that the robustness of minimum-phase and time-delay systems are derived analytically; and the Poisson integral formula is used to derive the robustness constraints of systems with right half plane (RHP) zero(s) and pole(s) [22-26]. It is shown that right half plane (RHP) zero(s) and/or time-delay of a plant limit the bandwidth of a DOB, however RHP pole(s) of a plant put(s) a lower bound on the bandwidth of a DOB to obtain a good robustness.

Besides that increasing the order of a DOB improves the performance of a system by using the bandwidth of a DOB more effectively; however the bandwidth constraints become more severe, and the robustness of a system deteriorates. New analysis and design methods are proposed by using the derived bandwidth constraints. The internal stability problem is solved by using an approximate minimum phase nominal plant model when a plant has non-minimum phase zero(s), and a performance controller is proposed for the conventional two-degrees-of-freedom control structure when a plant has RHP pole(s).

The rest of the paper is organized as follows. In section II, the conventional two-degrees-of-freedom control structure of a DOB based robust control system is presented briefly. In section III, the bandwidth constraints of a DOB are derived analytically by using the Bode and Poisson integral formulas. In section IV, four case-studies are given. The paper ends with conclusion given in the last section.

## 2. DISTURBANCE OBSERVER

Fig. 1 shows a general control block diagram for a DOB based robust control system. In this figure, $G(s)$ and $G_n(s)$ denote uncertain and nominal plant models, respectively; $Q(s)$ denotes the LPF of a DOB; $C(s)$ denotes the outer-loop controller; $r, \tau_{dis}$ and $\xi$ denote reference, disturbance and noise external inputs, respectively; and $\hat{\tau}_{dis}$ denotes estimated disturbances, which includes external disturbances and system uncertainties. The open-loop, sensitivity and co-sensitivity transfer functions are derived from the Fig. 1 as follows:

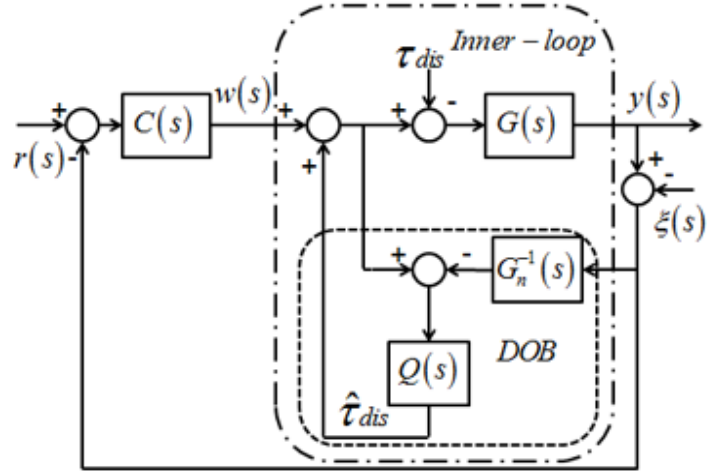

Fig. 1 A block diagram for a two-degrees-of-freedom DOB based robust control system

*Inner Loop:*

$$L_i(s) = \frac{G(s)Q(s)}{G_n(s)(1-Q(s))} \tag{1}$$

$$S_i = \frac{G_n(s)(1-Q(s))}{G_n(s)(1-Q(s)) + G(s)Q(s)} \tag{2}$$

$$T_i = \frac{G(s)Q(s)}{G_n(s)(1-Q(s)) + G(s)Q(s)} \tag{3}$$

*Outer Loop:*

$$L_o(s) = \frac{C(s)G(s)G_n(s)}{G_n(s)(1-Q(s)) + G(s)Q(s)} \tag{4}$$

$$S_o = \frac{G_n(s)(1-Q(s)) + G(s)Q(s)}{G_n(s)(1-Q(s)) + G(s)Q(s) + G(s)G_n(s)C(s)} \tag{5}$$

$$T_o = \frac{G(s)G_n(s)C(s)}{G_n(s)(1-Q(s)) + G(s)Q(s) + G(s)G_n(s)C(s)} \tag{6}$$

where $L_\bullet$, $S_\bullet$ and $T_\bullet$ denote the open-loop, sensitivity and co-sensitivity transfer functions, respectively [10].

**3. BANDWIDTH CONSTRAINTS OF DISTURBANCE OBSERVER**

In this section, the Bode and Poisson integral formulas are utilized to derive the bandwidth constraints of a DOB analytically. A multiplicative unstructured uncertainty is used to define the uncertain plant model given by

$$G(s) = G_n(s)(1 + \Delta W(s))\exp(-\tau s) \tag{7}$$

where $G(s)$ and $G_n(s)$ denote the uncertain and nominal plant models, respectively; $W(s)$ denotes the multiplicative unstructured uncertainty weighting function; and $\tau$ denotes delay time. Without losing the generality, a first order approximation of the weighting function is described by using

$$W(s) = \frac{w_T^{-1} s + e_{min}}{w_T^{-1} e_{max}^{-1} s + 1} \tag{8}$$

where $e_{min}$ and $e_{max}$ denote the minimum and maximum modeling errors, respectively; and $w_T$ is the frequency in which the nominal plant model starts to be a bad indicator for the uncertain plant [27,28]. It is assumed that $-e_{max}^{-1} < \Delta < 1$ instead of $|\Delta| < 1$ so that a RHP zero is not added due to uncertainty. The $n^{th}$ order LPF of a DOB is defined by using

$$Q(s) = \frac{g_0}{s^n + g_{n-1} s^{n-1} + \cdots + g_1 s + g_0} \tag{9}$$

If (1), (2) and (3) are re-written in terms of the LPF, plant uncertainty and time-delay, then

$$\begin{aligned} L_i(s) &= \frac{Q(s)}{(1-Q(s))}(1+\Delta W(s))e^{-\tau s} \\ S_i(s) &= \frac{(1-Q(s))}{(1-Q(s))+Q(s)(1+\Delta W(s))e^{-\tau s}} \\ T_i(s) &= \frac{Q(s)(1+\Delta W(s))e^{-\tau s}}{(1-Q(s))+Q(s)(1+\Delta W(s))e^{-\tau s}} \end{aligned} \tag{10}$$

The bandwidth constraints of a DOB are derived analytically as follows:

**3.1. Minimum-phase Plant**

*Lemma 1*: Let us consider the plant model given in (7) and assume that the uncertain plant is minimum-phase and the order of DOB is one. Then, it can be shown that the inner-loop is strictly robust if $\Delta > 0$, and its robustness can be guaranteed for a wide range of DOB's bandwidth if $\Delta < 0$. However, if a HODOB is used and/or the plant includes time delay, then the robustness of inner-loop cannot be guaranteed for a wide range of DOB's bandwidth even if $\Delta > 0$, and the bandwidth of a DOB becomes limited. ∎

*Proof:* Let us first consider a minimum-phase plant and re-write the open loop transfer function given in (10) by using a first order DOB.

$$L_i(s) = g_0 (1 + \Delta e_{max}) \frac{s + w_T e_{max}\left(\frac{1 + \Delta e_{min}}{1 + \Delta e_{max}}\right)}{s(s + w_T e_{max})} \qquad (11)$$

The equation (11) shows that the Nyquist plot of the inner-loop gets into the unit circle that is shown in Fig. 2 if $e_{min} > e_{max}$, which contradicts with the error assumption, when $\Delta > 0$. Thus, the first part of the *Lemma 1*, i.e., strict robustness, is satisfied.

Although the Nyquist plot of the inner-loop gets into the unit circle, i.e., strict roburtness is lost, when $\Delta < 0$, the robust stability can be guaranteed for a wide range of DOB's bandwidth as it can be seen from (11).

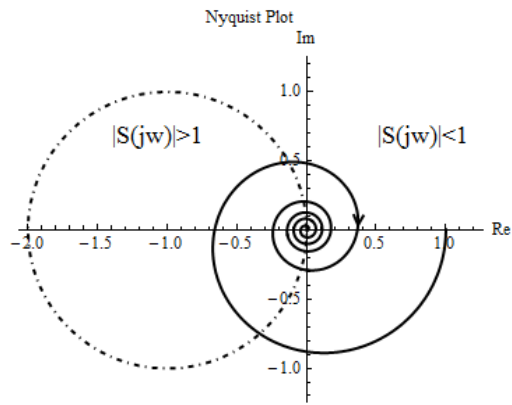

Fig. 2 Nyquist plot of inner-loop when a plant has time-delay and DOB is first order

However, if a HODOB is used instead of a DOB, then the robustness cannot be guaranteed for a wide range of DOB's bandwidth even if $\Delta > 0$. It can be easily shown by using a HODOB, e.g., a second order DOB. In that case, the robustness of a system can be guaranteed for a limited bandwidth of a HODOB.

The Fig. 2 shows the Nyquist plot of the inner-loop when a DOB is implemented to a time-delay plant. It indicates that the robustness cannot be guaranteed for a wide range of DOB's bandwidth if a plant includes time-delay. Hence, the proof of the *Lemma 1* is completed. ∎

The *Lemma 1* gives us a basic insight into the robustness of a DOB; however, further analysis is required for a HODOB. The Horowitzh's integral formula, given in (12), can be used to analyze the robustness of a HODOB [22,24].

$$\int_0^\infty \left( \log\left(|S_i(jw)|\right) - \log\left(|S_i(j\infty)|\right) \right) dw = -\frac{\pi}{2} \operatorname*{Res}_{s=\infty} \left( \log\left(S_i(s)\right) \right) \tag{12}$$

Since the relative degree of $L_i(s)$ is always higher than one when a HODOB is used, (12) can be simplified, and the Bode integral formula is obtained as follows: [22,24]

$$\int_0^\infty \log\left(|S_i(jw)|\right) dw = 0 \tag{13}$$

The robustness of a system depends on the magnitude of sensitivity function peak, which is defined by $\sup\left(|S(jw)|\right)$ where $\sup(*)$ denotes supremum of $*$. However, it is not an easy task to determine $\sup\left(|S(jw)|\right)$ by using (13) due to the infinite integral

range. Although, from a mathematical point of view, (13) can be balanced with a small peak in a wide frequency range, control systems cannot exhibit this response due to uncertainties, digital control implementations, and so on. Besides that the *Lemma 1* shows that $|S_i(jw)|$ has a peak if a HODOB is used. The *Lemma 2* bounds the integral range of (13) when a HODOB is used.

*Lemma 2:* Let us assume that $L_i(s)$ satisfies

$$|L_i(s)| \leq \frac{M}{w^{k+1}} = \delta \leq \frac{1}{2} \quad \forall w \geq w_\gamma \tag{14}$$

where $M \geq \limsup_{s \to \infty} |s^{k+1} L_i(s)|$ and $k+1$ is the order of DOB. Then, $S_i(s)$ satisfies

$$\left| \int_{w_\beta}^{\infty} \log(|S_i(jw)|) dw \right| \leq \frac{3\delta}{2k} w_\gamma \tag{15}$$

∎

*Proof:* The equation (14) holds if a HODOB is used. Let us consider the relation given by [30]

$$if \ |L_i(s)| \leq \frac{1}{2}, \ then \ |\log(1+L_i(s))| \leq \frac{3}{2}|L_i(s)| \leq \frac{3}{2}\delta \tag{16}$$

If (14) is put into (16), then

$$\left| \int_{w_\gamma}^{\infty} \log(|S_i(jw)|) dw \right| \leq \int_{w_\gamma}^{\infty} |\log(S_i(jw))| dw = \int_{w_\gamma}^{\infty} |\log(1+L_i(jw))| dw \leq \frac{3}{2} \int_{w_\gamma}^{\infty} \frac{M}{w^{1+k}} dw = \frac{3\delta}{2k} w_\gamma \tag{17}$$

∎

To derive the robustness constraints of a HODOB, the performance and robustness requirements are determined in a predefined frequency range by shaping the sensitivity transfer function. Then, the robustness of a system is analyzed, and the constraints are derived by using the *Theorem 1* as follows:

*Theorem 1:* Let us assume that a minimum phase plant is defined by using (7). Let us also assume that $S_i(s)$ satisfies $|S_i(jw)| \leq \alpha < 1, \forall w \leq w_\beta < w_\gamma$. If a DOB is used, then the system has a good robustness in a wide frequency range, yet its performance is limited by the dynamic characteristics of the DOB. However, if a HODOB is used, then the LPF of a HODOB should satisfy the following inequalities to obtain a good robustness and predefined performance criterion.

$$|Q(jw)| \geq \frac{1-\alpha}{1+\alpha|\Delta W(jw)|}, \forall w < \psi w_\gamma \quad \text{and} \quad \frac{|1-Q(j\psi w_\gamma)|}{|1+\Delta QW(j\psi w_\gamma)|} \geq \alpha \tag{18}$$

where $\psi = \dfrac{\sup\limits_{w\in[w_\beta,w_\gamma]} \log(|S_i(jw)|) + \dfrac{3\delta}{2k}}{\sup\limits_{w\in[w_\beta,w_\gamma]} \log(|S_i(jw)|) + \log(\alpha^{-1})}$ in which $|L_i(jw)| \leq \delta \leq \dfrac{1}{2}, \forall w \geq w_\gamma$. ∎

*Proof:* The *Lemma 1* proves the robustness of a DOB. Therefore, a HODOB can be considered directly. Let us rewrite (13) by using

$$\int_0^{w_\beta} \log(|S_i(jw)|)dw + \int_{w_\beta}^{w_\gamma} \log(|S_i(jw)|)dw + \int_{w_\gamma}^{\infty} \log(|S_i(jw)|)dw = 0 \tag{19}$$

If the sensitivity constraints given in the *Theorem 1* and *Lemma 2* are applied into (19), then

$$\log(\alpha)\int_0^{w_\beta} dw + \sup_{w\in[w_\beta,w_\gamma]}\log(|S_i(jw)|)\int_{w_\beta}^{w_\gamma} dw + \frac{3\delta}{2k}w_\gamma \geq 0 \tag{20}$$

which can be easily rewritten as follows:

$$\sup_{w\in[w_\beta,w_\gamma]} \log(|S_i(jw)|) \geq \log(\alpha^{-1})\frac{w_\beta}{w_\gamma - w_\beta} - \frac{3\delta}{2k}\frac{w_\gamma}{w_\gamma - w_\beta} \tag{21}$$

Journal of Dynamic Systems, Measurement and Control

$$\frac{w_\beta}{w_\gamma} \leq \psi = \frac{\sup_{w\in[w_\beta,w_\gamma]} \log(|S_i(jw)|) + \frac{3\delta}{2k}}{\sup_{w\in[w_\beta,w_\gamma]} \log(|S_i(jw)|) + \log(\alpha^{-1})} \quad (22)$$

where (22) is the function $\psi$ given in the *Theorem 1*. If the sensitivity constraint given in the *Theorem 1* is applied into (10), then

$$\frac{|1-Q(jw)|}{|1+\Delta QW(jw)|} \leq \alpha < 1, \ \forall w \leq w_\beta < w_\gamma \quad (23)$$

If (22) is applied into (23), then (18) is derived. ∎

As the order of a DOB, which is defined by $k+1$, is increased, the difference between the frequencies $w_\gamma$ and $w_\beta$ decreases. Therefore, increasing $k$ causes higher sensitivity peak as derived in (21).

The equation (23) shows that $\alpha^{-1}$ and $w_\beta$ increase as the bandwidth of a HODOB is increased. Therefore, the peak of $|S_i(jw)|$ becomes higher with the increasing bandwidth of a DOB as derived in (21).

The equation (18) provides a new design tool to obtain a good robustness and predefined performance criterion, which are determined by $\alpha$ and $w_\beta$. If the LPF of a HODOB satisfies (18), then the robustness and performance goals of a system can be achieved. However, (18) includes conservatism due to sectionally constant sensitivity bound defined by $|S_i(jw)| \leq \alpha < 1, \forall w \leq w_\beta < w_\gamma$. It can be lessened by using more realistic sensitivity bounds [29,30].

**3.2. Plant with Time-Delay**

The *Lemma 3* is used to bound the integral range of (13) when a plant has time-delay.

*Lemma 3:* Let us assume that $L_i(s)$ includes time-delay and satisfies

$$\left|L_i(s)\right| = \left|\tilde{L}_i(s) e^{-s\tau}\right| \leq \frac{M}{R^k} e^{-R\tau \cos(\theta)} \leq \delta \left(\frac{R}{|s|}\right)^k \quad \forall |s| \in S(R) \tag{24}$$

where $M \geq \limsup_{s \to \infty} \left|s^k L(s)\right|$; $k$ is the order of DOB; $s = R e^{j\theta}$; and $S(R) = \{s : \text{Re}(s) \geq 0 \text{ and } |s| \geq R\}$. Then, $S_i(s)$ satisfies

$$\left|\int_R^\infty \log\left(\left|S_i(jw)\right|\right) dw\right| \leq \frac{3\pi}{4\tau} \delta \tag{25}$$

∎

*Proof:* Similar to the *Lemma 2* [25].

∎

The bandwidth constraints of a DOB due to time-delay are derived by using the *Theorem 2* as follows:

*Theorem 2:* Let us assume that a plant is defined by using (7) where $G_n(s)$ is minimum-phase. Let us also assume that $S_i(s)$ satisfies $|S_i(jw)| \leq \alpha < 1$, $\forall w \leq w_\beta$. Then, the LPF of a DOB should satisfy the following inequalities to obtain a good robustness and predefined performance criterion.

$$\frac{|Q(jw)|}{|(1-Q(jw))|} \geq \frac{1-\alpha}{\alpha|1+\Delta W(jw)|}, \quad \forall w \leq \psi R \quad \text{and} \quad \frac{|1-Q(j\psi R)|}{|1-Q(j\psi R)+Q(j\psi R)(1+\Delta W(j\psi R))e^{-j\tau\psi R}|} \geq \alpha \tag{26}$$

If the order of a DOB is one, then

$$g \geq \frac{(1-\alpha)w}{\alpha|1+\Delta W(jw)|}, \quad \forall w \leq \psi R \quad \text{and} \quad \frac{\psi R}{|j\psi R + g(1+\Delta W(j\psi R))e^{-j\psi R\tau}|} \geq \alpha \tag{27}$$

where $\psi = \dfrac{\sup\limits_{w \in [w_\beta, R]} \log(|S_i(jw)|) + \dfrac{3\pi}{4\tau R}\delta}{\sup\limits_{w \in [w_\beta, R]} \log(|S_i(jw)|) + \log(\alpha^{-1})}$ in which $|L_i(s)| \leq \delta \left(\dfrac{R}{|s|}\right)^k \ \forall |s| \in S(R)$.

∎

*Proof:* Similar to the *Theorem 1*. The *Lemma 3* is used instead of the *Lemma 2*.

∎

The equation (26) provides a new design tool to obtain a good robustness and predefined performance criterion when a plant has time-delay. It shows that the bandwidth of a DOB is limited due to time-delay. However, the proposed design tool also includes conservatism, since the sensitivity bounds are not realistic.

Some comments are required to determine $R$. Its smallest value, which satisfies the constraint given in the *Theorem 2*, should be used to lessen the peak of $|S_i(jw)|$.

$$\sup_{s \in S(R)} (L_i(s)) = \max \left\{ \sup_{w \geq R} (|L_i(jw)|), \sup_{0 \leq \theta \leq \pi/2} (|L_i(Re^{j\theta})|) \right\} \quad (28)$$

The equation (28) shows that the *Theorem 2* holds even if $\sup\limits_{s \in S(R)}(L_i(s)) \geq \delta$, which can be used to lessen the peak of $|S_i(jw)|$ [25]. The sensitivity peak can be lessened if the following inequality holds.

$$\dfrac{w_T e_{\min}}{w^2 + (w_T e_{\min})^2} - \dfrac{w_T e_{\max}}{w^2 + (w_T e_{\max})^2} > \begin{cases} \tau, & 1^{st}\ order\ DOB \\ \tau + \dfrac{g_1}{s + g_1}, & 2^{nd}\ order\ DOB \\ \tau + \dfrac{g_1 g_2 + g_2 w^2}{g_2^2 w^2 + (g_1 - w^2)^2}, & 3^{rd}\ order\ DOB \\ \vdots \end{cases} \quad (29)$$

As it is expected from the *Theorem 1, (29)* shows that the peak of $|S_i(jw)|$ increases as the order of a DOB is increased.

### 3.3. Non-minimum Phase Plant

### *3.3.1. Performance Limitations*

It is a well-known fact that RHP zero(s) and pole(s) of open loop transfer functions cause undershoot and overshoot in the step responses of the closed loop systems, respectively. To achieve good performance, the following inequalities should be held.

$$w_B \leq \frac{2.1991 z_{RHP}}{\log\left(1 - \frac{0.9}{y_{undershoot}}\right)} \tag{30}$$

$$w_B \geq \frac{2.1991 p_{RHP}}{\log\left(10(y_{overshoot} - 0.9)\right)} \tag{31}$$

where $w_B, z_{RHP}, p_{RHP}, y_{undershoot}$ and $y_{overshoot}$, denote the bandwidth, RHP zero, RHP pole, infimum and supremum of the step response, respectively [27,28].

### *3.3.2. The Poisson's Integral Formulas*

The Poisson's integral formulas are used to derive the bandwidth constraints of a DOB analytically when a plant has RHP pole(s) and/or zero(s).

*Poisson's Integral Formulas:* Assume that an open-loop transfer function $L(s)$ has a RHP zero/(pole) at $z_{RHP} = \sigma_z + jw_z / (p_{RHP} = \sigma_p + jw_p)$. Let $S(s)/(T(s))$ be the sensitivity / (co-sensitivity) transfer function defined by $(1+L(s))^{-1} / (L(s)(1+L(s))^{-1})$. Then, it can be shown that the sensitivity / (co-sensitivity) transfer function satisfies

$$\int_{-\infty}^{\infty} \log(|S(jw)|) \frac{\sigma_z}{\sigma_z^2 + (w_z - w)^2} dw = \pi \log\left(|B_S^{-1}(z_{RHP})|\right) \tag{32}$$

$$\int_{-\infty}^{\infty} \log(|T(jw)|) \frac{\sigma_p}{\sigma_p^2 + (w_p - w)^2} dw = \pi \log\left(|B_T^{-1}(p_{RHP})|\right) \tag{33}$$

where $L(s) = \tilde{L}(s) B_S^{-1}(s) B_T(s)$; $\tilde{L}(s)$ is a minimum-phase transfer function; $B_S(s) = \prod_{i=1}^{k} \frac{p_i - s}{\overline{p}_i + s}$ and $B_T(s) = \prod_{i=1}^{l} \frac{z_i - s}{\overline{z}_i + s}$ are Blaschke products [29,30]. The integral ranges of (32) and (33) are bounded by $W_{pi}(w) = \frac{\sigma_x}{\sigma_x^2 + (w_x - w)^2}$.

### 3.3.3 Plant with RHP Zero(s)

An approximate nominal plant model is used to solve the internal stability problem when a plant has RHP zero(s). It is defined by

$$G(s) = G_n(s)(1 + \Delta W_T(s)) = \hat{G}_n(s) r_{err}(s)(1 + \Delta W_T(s)) \tag{34}$$

where $\hat{G}_n(s)$ is the approximate nominal plant model, which has stable inverse; and $r_{err}(s) = G_n(s)(\hat{G}_n(s))^{-1}$ [31]. Then, the open-loop transfer functions are defined as follows

*Inner Loop:* $$L_i(s) = \frac{r_{err}(s)(1 + \Delta W_T(s)) Q(s)}{1 - Q(s)} \tag{35}$$

*Outer Loop:* $$L_o(s) = \frac{C(s) G(s)}{1 - Q(s) + r_{err}(s)(1 + \Delta W_T(s)) Q(s)} \tag{36}$$

The bandwidth constraints of a DOB due to RHP zero(s) are derived by using the *Theorem 3* as follows:

*Theorem 3:* Let us assume that a plant, which has a RHP zero at $z_{RHP}$, is defined by using (7) when $\tau = 0$; and $S_i(s)$ and $T_i(s)$ are the sensitivity and co-sensitivity transfer functions of the inner-loop, respectively. Let us also assume that the frequency responses of the sensitivity and co-sensitivity transfer functions satisfy $|S_i(jw)| \leq \alpha_\beta, \forall w \leq w_\beta$ and $|T_i(jw)| \leq \alpha_\gamma, \forall w \geq w_\gamma$. Then, the LPF of a DOB should satisfy the

following constraints to obtain a good robustness and predefined performance criteria.

$$\frac{|Q(jw)|}{|1-Q(jw)|} \geq \frac{1-\alpha_\beta}{\alpha_\beta |r_{er}(jw)(1+\Delta W_T(jw))|}, \forall w < z_{RHP}, \psi_1 \text{ and } \frac{|1-Q(jz_{RHP}\psi_1)|}{|1-Q(jz_{RHP}\psi_1)+r_{er}Q(jz_{RHP}\psi_1)(1+\Delta W_T(jz_{RHP}\psi_1))|} \geq \alpha_\beta \quad (37)$$

$$\frac{|r_{er}Q(jw)(1+\Delta W_T(jw))|}{|1-Q(jw)|} \leq \frac{\alpha_\gamma}{1-\alpha_\gamma}, \forall w > z_{RHP}, \psi_2 \text{ and } \frac{|r_{er}Q(jz_{RHP}\psi_2)(1+\Delta W_T(jz_{RHP}\psi_2))|}{|1-Q(jz_{RHP}\psi_2)+r_{er}Q(jz_{RHP}\psi_2)(1+\Delta W_T(jz_{RHP}\psi_2))|} \geq \alpha_\gamma \quad (38)$$

where 
$$\psi_1 = \tan\left(\frac{\log(1+\alpha_\gamma)(\pi-2\vartheta(w_\gamma))+2\log\left(\max_{w_\beta \leq w \leq w_\gamma}(|S(jw)|)\right)\vartheta(w_\gamma)}{2\left(\log\left(\max_{w_\beta \leq w \leq w_\gamma}(|S(jw)|)\right)+\log(\alpha_\beta^{-1})\right)} - \frac{\pi\log(|B_S^{-1}(z_{RHP})|)}{2\left(\log\left(\max_{w_\beta \leq w \leq w_\gamma}(|S(jw)|)\right)+\log(\alpha_\beta^{-1})\right)}\right),$$

$$\psi_2 = \tan\left(\frac{\log\left((1+\alpha_\gamma)^{-1}\right)\pi+2\left(\log(\alpha_\beta^{-1})+\log\left(\max_{w_\beta \leq w \leq w_\gamma}(|S(jw)|)\right)\right)\vartheta(w_\beta)}{2\left(\log\left(\max_{w_\beta \leq w \leq w_\gamma}(|S(jw)|)\right)+\log\left((1+\alpha_\gamma)^{-1}\right)\right)} + \frac{\pi\log(|B_S^{-1}(z_{RHP})|)}{2\left(\log\left(\max_{w_\beta \leq w \leq w_\gamma}(|S(jw)|)\right)+\log(\alpha_\beta^{-1})\right)}\right),$$

$$\vartheta(w_\bullet) = \int_0^{w_\bullet} W dw = \int_0^{w_\bullet} \frac{z_{RHP}}{z_{RHP}^2 + w^2} dw = \arctan\left(\frac{w_\bullet}{z_{RHP}}\right) \qquad \blacksquare$$

*Proof:* If $|T_i(jw)| \leq \alpha_\gamma, \forall w \geq w_\gamma$, then $|S_i(jw)| \leq 1+\alpha_\gamma, \forall w \geq w_\gamma$. If the sensitivity constraints are applied into (32), then

$$\log(1+\alpha_\gamma)\left(\int_{-\infty}^{-w_\gamma} W(w)dw + \int_{w_\gamma}^{\infty} W(w)dw\right) + \log(\alpha_\beta)\int_{-w_\beta}^{w_\beta} W(w)dw + \log\left(\max_{w_\beta \leq w \leq w_\gamma}(|S_i(jw)|)\right)\left(\int_{-w_\gamma}^{-w_\beta} W(w)dw + \int_{w_\beta}^{w_\gamma} W(w)dw\right) \geq \pi\log(|B_S^{-1}(z_{RHP})|) \quad (39)$$

which can be transformed into

$$\log\left(\max_{w_\beta \leq w \leq w_\gamma}(|S_i(jw)|)\right) \geq \log(\alpha_\beta^{-1})\frac{2\vartheta(w_\beta)}{2(\vartheta(w_\gamma)-\vartheta(w_\beta))} + \log\left((1+\alpha_\gamma)^{-1}\right)\frac{\pi-2\vartheta(w_\gamma)}{2(\vartheta(w_\gamma)-\vartheta(w_\beta))} + |B_S^{-1}(z_{RHP})|\frac{\pi}{2(\vartheta(w_\gamma)-\vartheta(w_\beta))} \quad (40)$$

$$\vartheta(w_\beta) \leq \frac{\log(1+\alpha_\gamma)(\pi-2\vartheta(w_\gamma))+2\log\left(\max_{w_\beta \leq w \leq w_\gamma}(|S_i(jw)|)\right)\vartheta(w_\gamma)}{2\left(\log\left(\max_{w_\beta \leq w \leq w_\gamma}(|S_i(jw)|)\right)+\log(\alpha_\beta^{-1})\right)} - \frac{\pi\log(|B_S^{-1}(z_{RHP})|)}{2\left(\log\left(\max_{w_\beta \leq w \leq w_\gamma}(|S_i(jw)|)\right)+\log(\alpha_\beta^{-1})\right)} \quad (41)$$

$$\vartheta(w_\gamma) \geq \frac{\log\left((1+\alpha_\gamma)^{-1}\right)\pi + 2\left(\log(\alpha_\beta^{-1}) + \log\left(\max_{w_\beta \leq w \leq w_\gamma}(|S_i(jw)|)\right)\right)\vartheta(w_\beta)}{2\left(\log\left(\max_{w_\beta \leq w \leq w_\gamma}(|S_i(jw)|)\right) + \log\left((1+\alpha_\gamma)^{-1}\right)\right)} + \frac{\pi \log\left(|B_S^{-1}(z_{RHP})|\right)}{2\left(\log\left(\max_{w_\beta \leq w \leq w_\gamma}(|S_i(jw)|)\right) + \log(\alpha_\beta^{-1})\right)} \quad (42)$$

where (41) and (42) are the functions $\psi_1$ and $\psi_2$ given in *the Theorem 3*. If the sensitivity and co-sensitivity transfer functions are derived by using (35), and their constraints given in the *Theorem 3* are applied, then

$$\left| \frac{1-Q(s)}{1-Q(s)+r_{err}(s)(1+\Delta W_T(s))Q(s)} \right| \leq \alpha_\beta, \forall w \leq w_\beta \quad (43)$$

$$\left| \frac{r_{err}(s)(1+\Delta W_T(s))Q(s)}{1-Q(s)+r_{err}(s)(1+\Delta W_T(s))Q(s)} \right| \leq \alpha_\gamma, \forall w \geq w_\gamma \quad (44)$$

If (41) and (42) are applied into (43) and (44), then (37) and (38) are derived directly. ∎

The equations (37) and (38) provide new design tools to obtain good robustness and predefined performance criteria when a plant has RHP zero(s). They show us that the bandwidth of a DOB is limited due to RHP zero(s). The performance at high frequencies is controlled by bounding the co-sensitivity transfer function, so the bounds of the DOB's bandwidth are improved. However, the *Theorem 3* still includes conservatism due to the unrealistic bounds of sensitivity and co-sensitivity transfer functions.

### 3.3.4. Plant with RHP Pole(s)

Against the conventional design of DOB based robust control systems, the outer loop controller should be designed as a stabilizing controller instead of a performance

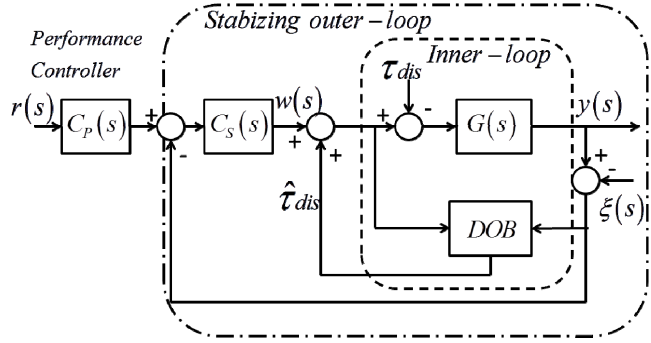

Fig. 3. A block diagram of a DOB based robust control system when a plant is unstable

one, since the inner-loop is always unstable when a plant has RHP pole(s). Therefore, a new controller $C_P(s)$, shown in Fig. 3, is used to achieve performance requirements. The bandwidth constraint due to RHP pole(s) is derived by using the *Theorem 4* as follows:

*Theorem 4:* Let us assume that a plant, which has a RHP pole at $p_{RHP}$, is defined by using (7) when $\tau = 0$. Let us also assume that the nominal plant can be stabilized by using the stabilizing outer-loop controller $C_S(s)$, and the co-sensitivity function of the outer-loop $T_o(s)$ satisfies $|T_o(jw)| \leq \alpha, \forall w \geq w_\alpha$. Then, the LPF of a DOB should satisfy the following constraint to obtain a good robustness and predefined performance criterion.

$$|1+\Delta Q(jw)W_T(jw)| \geq \frac{1-\alpha}{\alpha}|C_s(jw)G(jw)|, \quad \forall w \geq p_{RHP}\psi \qquad (45)$$

where $\psi = \tan\left(\dfrac{\pi\left(\log(\alpha^{-1})+\log(|B_T^{-1}(p_{RHP})|)\right)}{2\left(\log(\alpha^{-1})+\log(\|T_o\|_\infty)\right)}\right)$.  ∎

*Proof:* Similar to the *Theorem 3*.  ∎

The equation (45) provides a new design tool when a plant has RHP pole(s). It shows that the bandwidth of a DOB has a lower bound to obtain a good robustness when the plant is unstable; however, it also suffers from the conservatism.

Consequently, the design constraints of a DOB are summarized in *Table 1*.

Table 1: Summary of the design constraints of a DOB

| | | |
|---|---|---|
| Minimum Phase Systems | First Order DOB | Good Robustness. Limited Performance. |
| | Higher Order DOB | $\|Q(jw)\| \geq \dfrac{1-\alpha}{1+\alpha\|\Delta W(jw)\|}, \forall w < \psi w_\gamma$ and $\dfrac{\|1-Q(j\psi w_\gamma)\|}{\|1+\Delta QW(j\psi w_\gamma)\|} \geq \alpha$ (18) |
| Plant with Time Delay | | $\dfrac{\|Q(jw)\|}{\|(1-Q(jw))\|} \geq \dfrac{1-\alpha}{\alpha\|1+\Delta W(jw)\|}, \forall w \leq \psi R$ and $\dfrac{\|1-Q(j\psi R)\|}{\|1-Q(j\psi R)+Q(j\psi R)(1+\Delta W(j\psi R))e^{-j\tau\psi R}\|} \geq \alpha$ (26) |
| Plant with RHP Zero | | $\dfrac{\|Q(jw)\|}{\|1-Q(jw)\|} \geq \dfrac{1-\alpha_\beta}{\alpha_\beta\|r_{er}(jw)(1+\Delta W_T(jw))\|}, \forall w < z_{RHP}\psi_1$ and $\dfrac{\|1-Q(jz_{RHP}\psi_1)\|}{\|1-Q(jz_{RHP}\psi_1)+r_{er}Q(jz_{RHP}\psi_1)(1+\Delta W_T(jz_{RHP}\psi_1))\|} \geq \alpha_\beta$ (37) |
| | | $\dfrac{\|r_{er}Q(jw)(1+\Delta W_T(jw))\|}{\|1-Q(jw)\|} \leq \dfrac{\alpha_\gamma}{1-\alpha_\gamma}, \forall w > z_{RHP}\psi_2$ and $\dfrac{\|r_{er}Q(jz_{RHP}\psi_2)(1+\Delta W_T(jz_{RHP}\psi_2))\|}{\|1-Q(jz_{RHP}\psi_2)+r_{er}Q(jz_{RHP}\psi_2)(1+\Delta W_T(jz_{RHP}\psi_2))\|} \geq \alpha_\gamma$ (38) |
| Plant with RHP Pole | | $\|1+\Delta Q(jw)W_T(jw)\| \geq \dfrac{1-\alpha}{\alpha}\|C_S(jw)G(jw)\|, \forall w \geq p_{RHP}\psi$ (45) |

## 4. SIMULATION RESULTS

In this section, four case studies are carried out to verify the proposals.

### 4.1. Minimum-Phase Plant:

Let us consider nominal and uncertain plant models by using

$$G_n(s) = \frac{s+5}{s^2+5s+6} \text{ and } G(s) = G_n(s)(1+\Delta W_T(s)) \quad (46)$$

where $W_T(s) = \dfrac{5s+100}{s+500}$ and $-0.2 < \Delta < 1$. The bandwidth of DOB should be smaller than $100\,rad/s.$ to obtain robust stability. Fig. 4 shows that as the order of DOB is increased, the bandwidth of DOB is used more effectively and noise suppression is

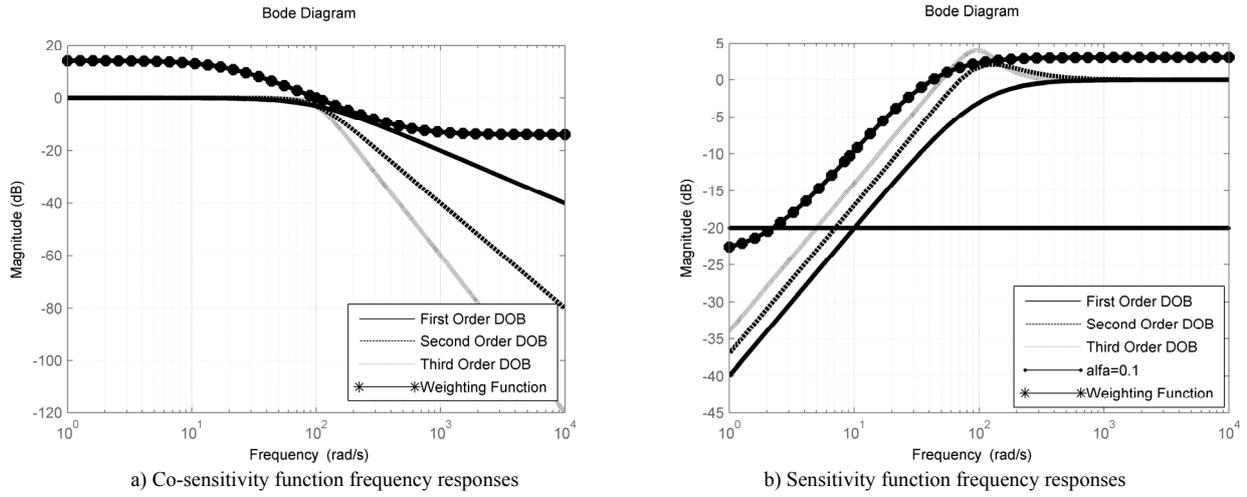

a) Co-sensitivity function frequency responses  b) Sensitivity function frequency responses

Fig. 4 Frequency responses of the inner-loop sensitivity and co-sensitivity transfer functions when the bandwidth of DOB is 100 rad/s.

improved. However, the robustness deteriorates, and the bandwidth constraints of DOB become more severe, as expected from the *Theorem 1.* The robustness and performance of the system can be improved by using the *Theorem 1* as follows:

The design parameters $k$, $\alpha$ and $\sup_{w \in [w_\beta, R]} \log(|S(jw)|)$ should be determined by considering the robustness and performance design criteria. The order of DOB determines $\delta$, which is directly related to the slope of $L_i(s)$. If a second order DOB is used, and the design parameters $\alpha$ and $\sup_{w \in [w_\beta, w_r]} \log(|S(jw)|)$ are chosen as $0.1$ and $\sqrt{2}$, respectively, then $\delta = 0.4$ satisfies for $\forall w \geq 100 \, rad/s.$. Hence, the performance and robustness constraints are derived by using (18) and (21) as follows:

$$w_\beta \leq 46 \, rad/s. \text{ and } B_W < 100 \, rad/s. \tag{47}$$

where $B_W$ denotes the bandwidth of DOB. Fig. 5 shows the frequency responses of the inner-loop sensitivity and co-sensitivity transfer functions when a second order DOB has different bandwidth values. The performance and robustness constraints, which are different from (47) due to conservatism, are obtained directly from the Fig. 5 as follows:

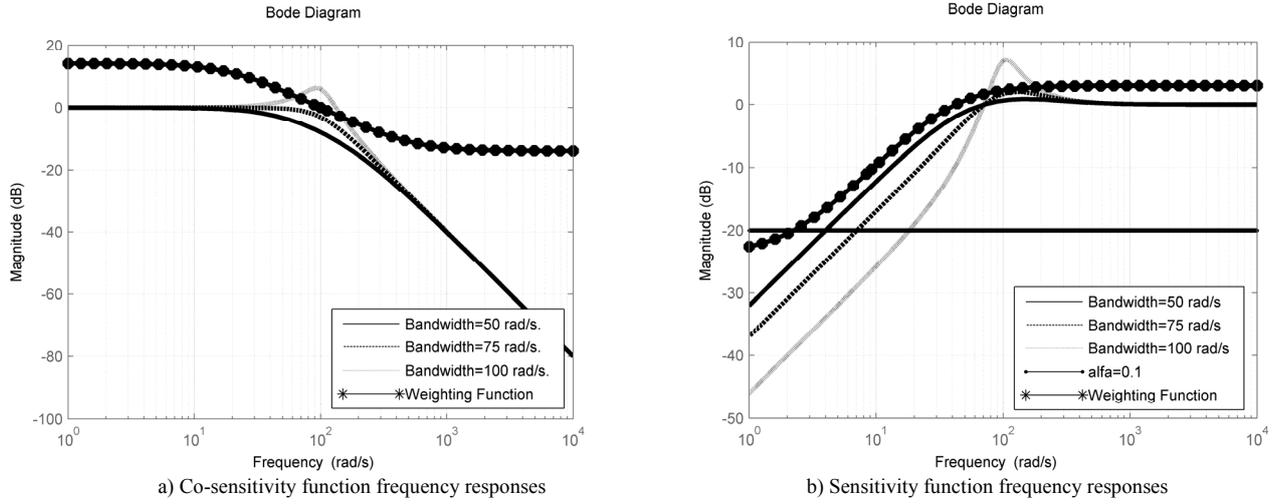

a) Co-sensitivity function frequency responses  b) Sensitivity function frequency responses

Fig. 5 Frequency responses of the inner-loop sensitivity and co-sensitivity transfer functions when a 2$^{nd}$ order DOB is used.

$$w_\beta \leq 15\,rad/s.\ and\ B_W < 65\,rad/s. \tag{48}$$

The weighting function of the sensitivity transfer function is $W_S(s) = \dfrac{0.707s+30}{s+2}$.

## 4.2 Plant with Time-Delay

Let us consider the time-delay constraints by using the following plant model.

$$G_n(s) = \frac{s+10}{s^2+5s+10}\ and\ G(s) = G_n(s)(1+\Delta W_T(s))e^{-0.01s} \tag{49}$$

where $W_T(s) = \dfrac{3s+240}{s+600}$. The bandwidth constraint of DOB is obtained by using the

*Theorem 2* as follows:

The design parameters $\alpha$ and $\sup_{w\in[w_\beta,R]} \log(|S_i(jw)|)$ should be determined by considering the performance and robustness design criteria. Besides, $\delta$ and $R$, which depend on the order of DOB, should be determined. Let us assume that $\alpha = 0.1$, $\sup_{w\in[w_\beta,R]} \log(|S_i(jw)|) = 2$ and the order of DOB is one. Since a first order DOB is used, if we take $\delta = 0.1$, then $R \cong B_W \delta^{-1} = 10 B_W$, where $B_W$ is the bandwidth of DOB. If (26) is used,

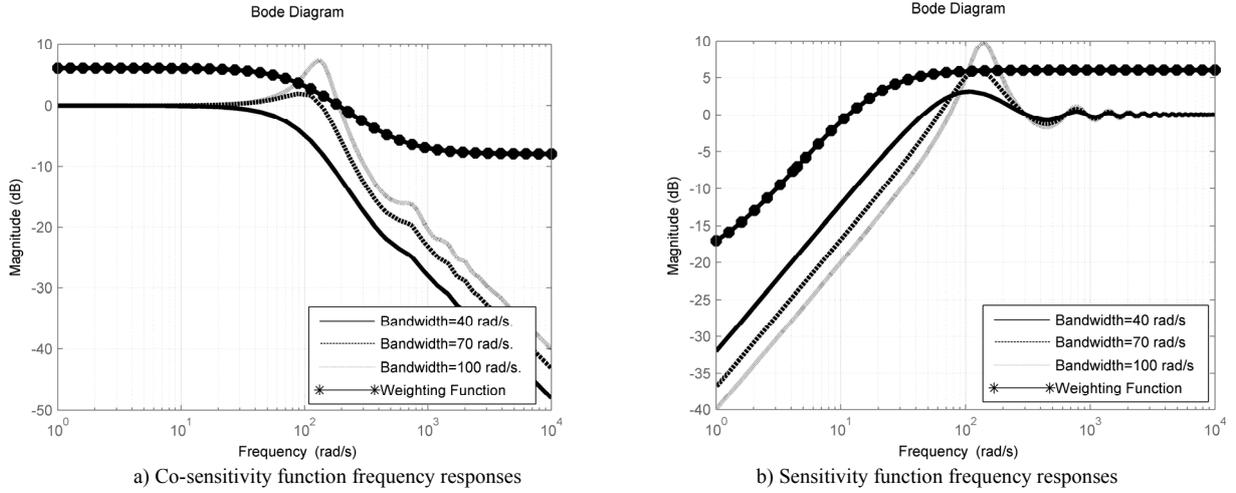

a) Co-sensitivity function frequency responses  b) Sensitivity function frequency responses

Fig. 6. Frequency responses of the inner-loop sensitivity and co-sensitivity transfer functions when the order of DOB is one.

then $\sup_{w \in [w_\beta, R]} \log(|S_i(jw)|) \leq 2$ is satisfied for a wide range of DOB's bandwidth. If a second order DOB is used, then $\sup_{w \in [w_\beta, R]} \log(|S(jw)|) \leq \sqrt{2}$ is satisfied for $B_W \leq 500 \, rad/s.$. Fig. 6 shows the frequency responses of inner-loop sensitivity and co-sensitivity transfer functions. The real bandwidth constraint, which is significantly different from the derived one due to conservatism, is obtained as $B_W \leq 70 \, rad/s.$. The sensitivity weighting function is same as given above.

### 4.3. Plant with a RHP Zero

Let us consider the RHP zero constraints by using the following plant model.

$$G_n(s) = \frac{-s+50}{s^2+25s+40} \text{ and } G(s) = G_n(s)(1+\Delta W_T(s)) \tag{50}$$

where $W_T(s) = \frac{3.75s+450}{s+1500}$ and $\hat{G}_n \cong \frac{s^2+200s+20}{(4s+0.4)(s^2+25s+40)}$ [31]. It is assumed that $\alpha_\beta = 0.5$, $\alpha_\gamma = 0.2$, $\|S\|_\infty = 2$ and $w_\gamma = 2B_W$. The performance and robustness constraints are derived by using the *Theorem 3* as follows:

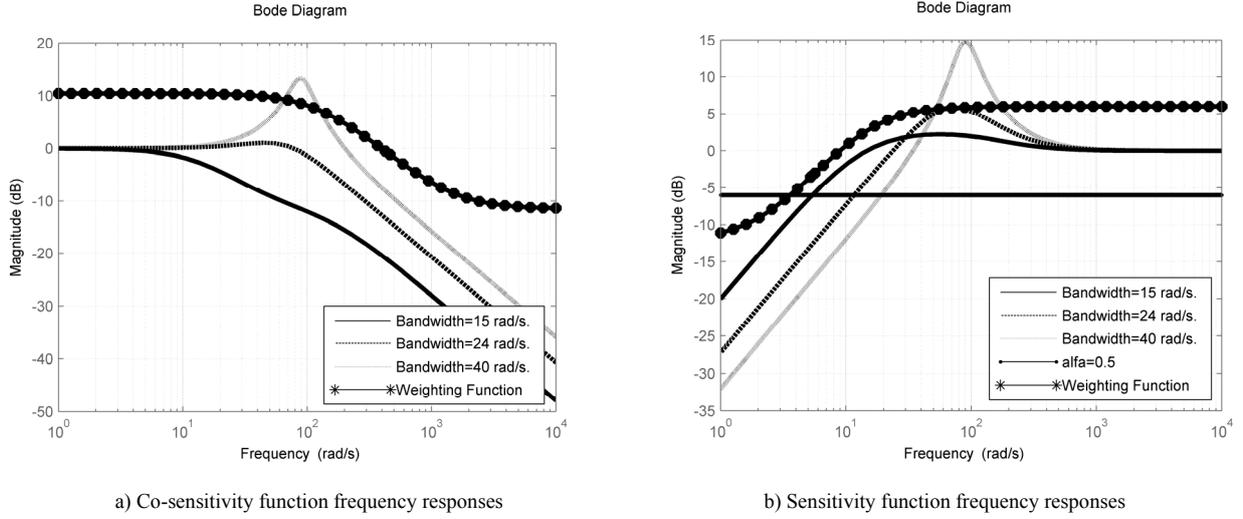

a) Co-sensitivity function frequency responses  b) Sensitivity function frequency responses

Fig. 7. Frequency responses of the inner-loop sensitivity and co-sensitivity transfer functions when the order of DOB is one.

$$6 \leq B_W \leq 55, \ w_\beta \leq 35 \, rad/\sec \qquad (51)$$

Fig. 7 shows the frequency responses of inner-loop sensitivity and co-sensitivity transfer functions when a first order DOB is used. The performance and bandwidth constraints, which are different from (51) due to conservatism, are obtained directly from the Fig.7 as follows:

$$12 \leq BW \leq 24 \text{ and } w_\beta \leq 15 \, rad/\sec \qquad (52)$$

The sensitivity weighting function and the outer loop controller, which satisfies (30), are used given by $W_S(s) = 0.5 \frac{s+16}{s+2}$ and $C(s) = 1 + 0.1s + \frac{4}{s}$, respectively.

### 4.4. Plant with a RHP Pole:

Let us consider the RHP pole constraints by using the following plant model.

$$G_n(s) = \frac{1}{s(s-5)} \text{ and } G(s) = G_n(s)(1 + \Delta W_T(s)) \qquad (53)$$

where $W_T(s) = \frac{7.5s + 600}{s + 1500}$. Since the DOB is implemented in the inner-loop, the stabilizing

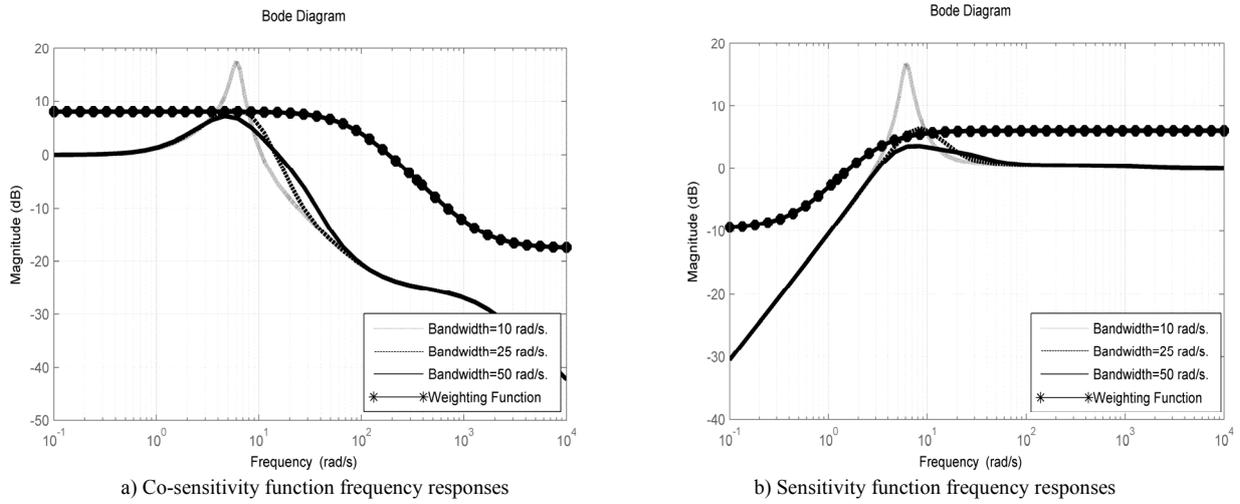

a) Co-sensitivity function frequency responses    b) Sensitivity function frequency responses

Fig. 8. Frequency responses of the outer-loop sensitivity and co-sensitivity transfer functions when the order of DOB is two

outer loop controller, $C_S(s) = 20 + 12s$, is designed by considering only the nominal plant model, and the robustness is achieved by using low control signals. The frequency responses of outer-loop sensitivity and co-sensitivity transfer functions can be seen in Fig. 8 when a second order DOB is used. It clearly shows that the robustness of the system improves as the bandwidth of DOB is increased. The lower bound on the bandwidth of DOB can be derived by using the *Theorem 4* similarly; however it also includes conservatism.

Fig. 9a shows the performance improvement of the controller $C_P(s)$ which is designed as $C_P(s) = \dfrac{s+10}{4s+10}$. Fig 9b shows the step responses for different external disturbances. As it can be seen from the figure, the DOB cannot estimate high frequency disturbances precisely, so the robustness of the system deteriorates. There is a trade-off between the robustness and noise response to determine the bandwidth of DOB.

**4.5. Comments on Conservatism:**

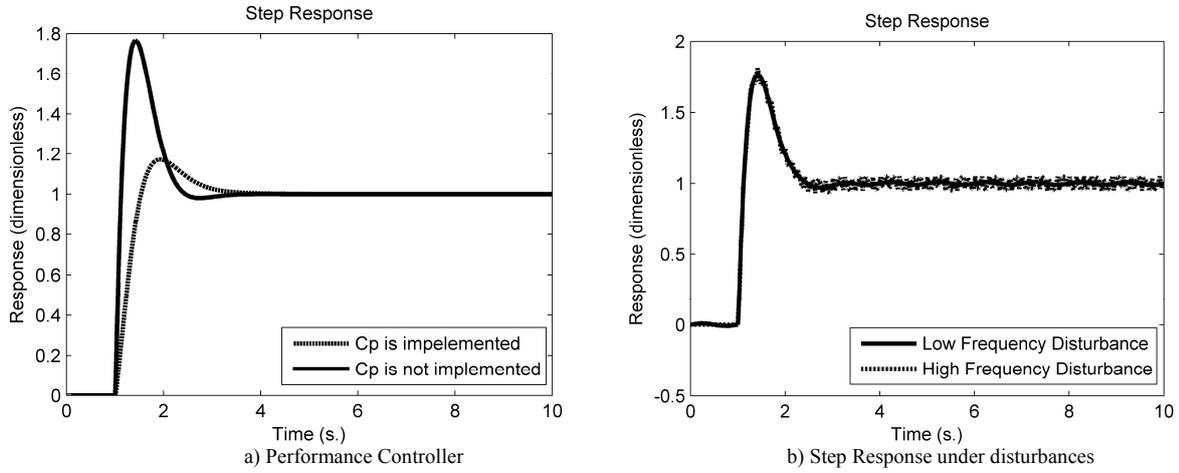

a) Performance Controller        b) Step Response under disturbances

Fig. 9. Step responses of the unstable plant when it is controlled by using the proposed robust controller

The source of conservatism, i.e., the approximate sensitivity/co-sensitivity bounds, can be seen in Fig. 10. In this figure, the grey areas are determined by the sectionally constant sensitivity bounds, and the black areas denote the errors which cannot be considered in the robustness analysis. It clearly shows that, there is a significant difference between the areas, which are bounded by the real sensitivity function and its approximate bound.

Let us again consider the plant with time-delay example, which has the most severe conservative result, to obtain more accurate bandwidth constraint by decreasing the conservatism. If we consider the nominal plant model with time-delay, then the sensitivity function frequency response is derived as follows:

$$|S(jw)|^2 = \frac{w^2}{w^2 - 2gw\sin(w\tau) + g^2} \quad (54)$$

where $g$ is the bandwidth of the first order DOB; and $\tau$ is delay time. The frequencies $w_1$ and $w_2$, given in the Fig. 10, can be obtained approximately as follows:

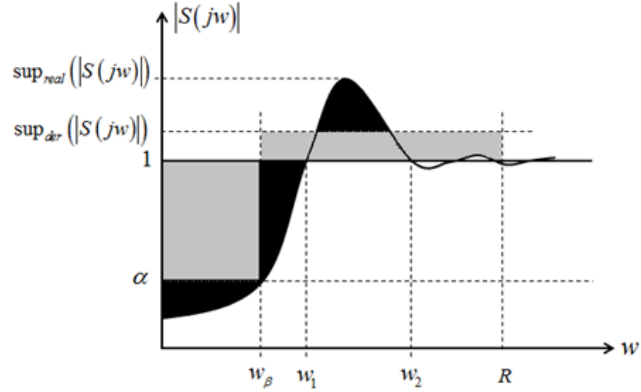

Fig. 10 A general frequency response of a sensitivity function

$$w_1 = \sqrt{\frac{3\tau - 1.73205\sqrt{\tau^2(3-g_0\tau)}}{\tau^3}}, \quad w_2 = \sqrt{\frac{3\tau + 1.73205\sqrt{\tau^2(3-g_0\tau)}}{\tau^3}} \tag{55}$$

The conservatism can be decreased by using $w_1$ and $w_2$ instead of $R$. If we use the sectionally constant sensitivity bound with $w_1$ and $w_2$, then the bandwidth limitation of DOB, which is obtained as 70 rad/s in the second example, is derived as 95 rad/s. Hence, the conservatism can be lessened by considering more realistic sensitivity bounds.

## 5. CONCLUSION

This paper has been concerned with the problem of robustness and performance constraints in the design of DOB based control systems. The bandwidth constraints of DOB are derived analytically by using the Bode and Poisson integral formulas, and new analysis and design tools are proposed. The proposed tools include conservatism; however, it can be lessened by using a more realistic sensitivity/co-sensitivity bound, which increases the complexity of analysis, as shown in the paper. The experiences with the proposed tools showed us that the conservatism is not a severe problem, since the proposed tools give a deep insight into the design constraints of DOB based control

Journal of Dynamic Systems, Measurement and Control

systems. Therefore, the analysis and design tools are very useful, and they can be easily implemented into many different DOB based robust control problems.

**ACKNOWLEDGMENT**

This research was supported in part by the Ministry of Education, Culture, Sports, Science and Technology of Japan under Grant-in-Aid for Scientific Research (S), 25220903, 2013.

**Table Caption List**

Table 1     Summary of the design constraints of a DOB

# Figure Captions List

Table 1: Summary of the design constraints of a DOB

| | | |
|---|---|---|
| Minimum Phase Systems | First Order DOB | Good Robustness. Limited Performance. |
| | Higher Order DOB | $\|Q(jw)\| \geq \dfrac{1-\alpha}{1+\alpha\|\Delta W(jw)\|}, \forall w < \psi w_\gamma$ and $\dfrac{\|1-Q(j\psi w_\gamma)\|}{\|1+\Delta QW(j\psi w_\gamma)\|} \geq \alpha$ (18) |
| Plant with Time Delay | | $\dfrac{\|Q(jw)\|}{\|(1-Q(jw))\|} \geq \dfrac{1-\alpha}{\alpha\|1+\Delta W(jw)\|}, \forall w \leq \psi R$ and $\dfrac{\|1-Q(j\psi R)\|}{\|1-Q(j\psi R)+Q(j\psi R)(1+\Delta W(j\psi R))e^{-j\tau\psi R}\|} \geq \alpha$ (26) |
| Plant with RHP Zero | | $\dfrac{\|Q(jw)\|}{\|1-Q(jw)\|} \geq \dfrac{1-\alpha_\beta}{\alpha_\beta\|r_{err}(jw)(1+\Delta W_T(jw))\|}, \forall w < z_{RHP}\psi_1$ and $\dfrac{\|1-Q(jz_{RHP}\psi_1)\|}{\|1-Q(jz_{RHP}\psi_1)+r_{err}Q(jz_{RHP}\psi_1)(1+\Delta W_T(jz_{RHP}\psi_1))\|} \geq \alpha_\beta$ (37) <br><br> $\dfrac{\|r_{err}Q(jw)(1+\Delta W_T(jw))\|}{\|1-Q(jw)\|} \leq \dfrac{\alpha_\gamma}{1-\alpha_\gamma}, \forall w > z_{RHP}\psi_2$ and $\dfrac{\|r_{err}Q(jz_{RHP}\psi_2)(1+\Delta W_T(jz_{RHP}\psi_2))\|}{\|1-Q(jz_{RHP}\psi_2)+r_{err}Q(jz_{RHP}\psi_2)(1+\Delta W_T(jz_{RHP}\psi_2))\|} \geq \alpha_\gamma$ (38) |
| Plant with RHP Pole | | $\|1+\Delta Q(jw)W_T(jw)\| \geq \dfrac{1-\alpha}{\alpha}\|C_S(jw)G(jw)\|, \quad \forall w \geq p_{RHP}\psi$ (45) |

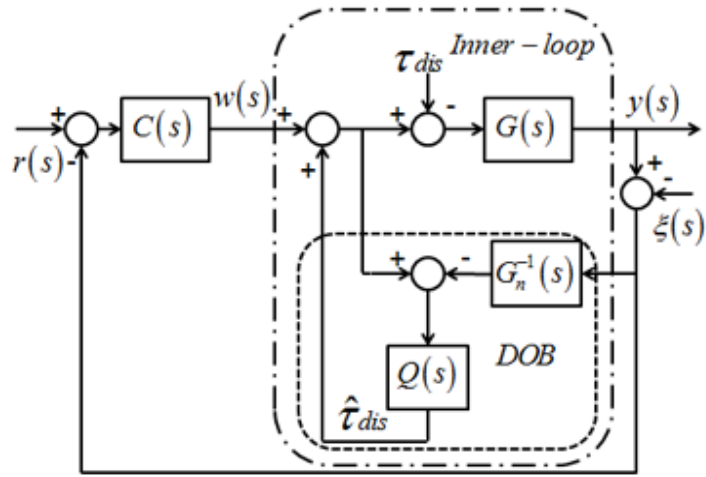

Fig. 1 A block diagram for a two-degrees-of-freedom DOB based robust control system

Journal of Dynamic Systems, Measurement and Control

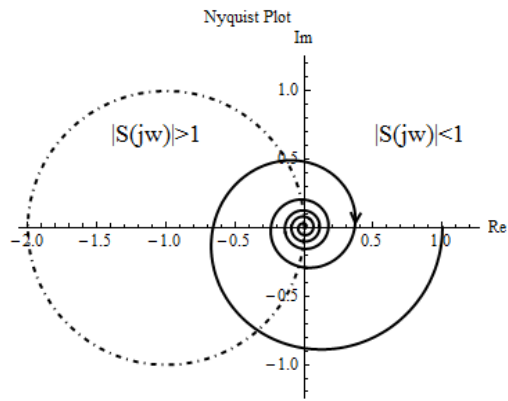

Fig. 2 Nyquist plot of inner-loop when a plant has time-delay and DOB is first order

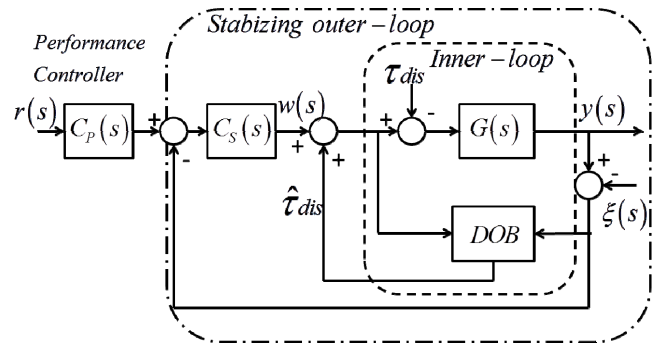

Fig. 3.A block diagram of a DOB based robust control system when a plant is unstable

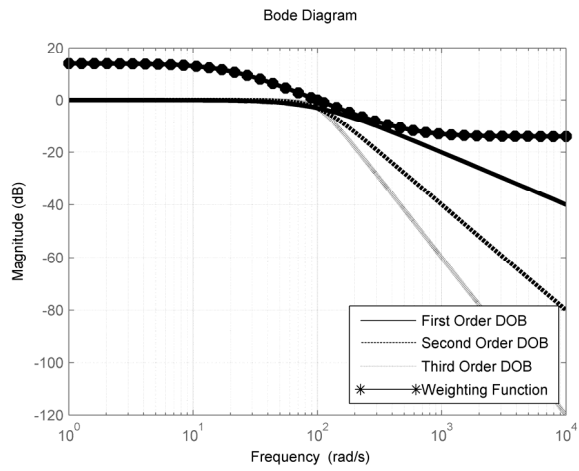
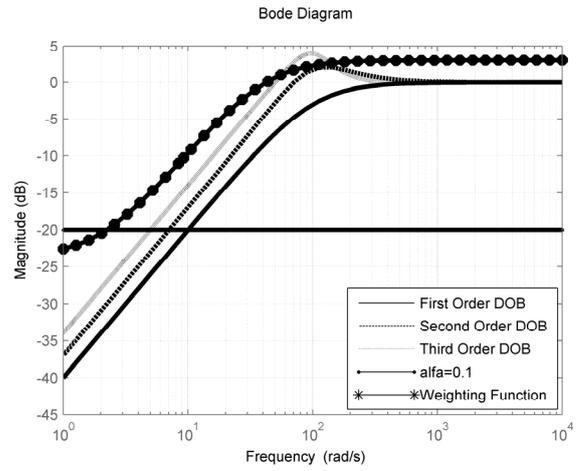

a) Co-sensitivity function frequency responses     b) Sensitivity function frequency responses

Fig. 4 Frequency responses of the inner-loop sensitivity and co-sensitivity transfer functions when the bandwidth of DOB is 100 rad/s.

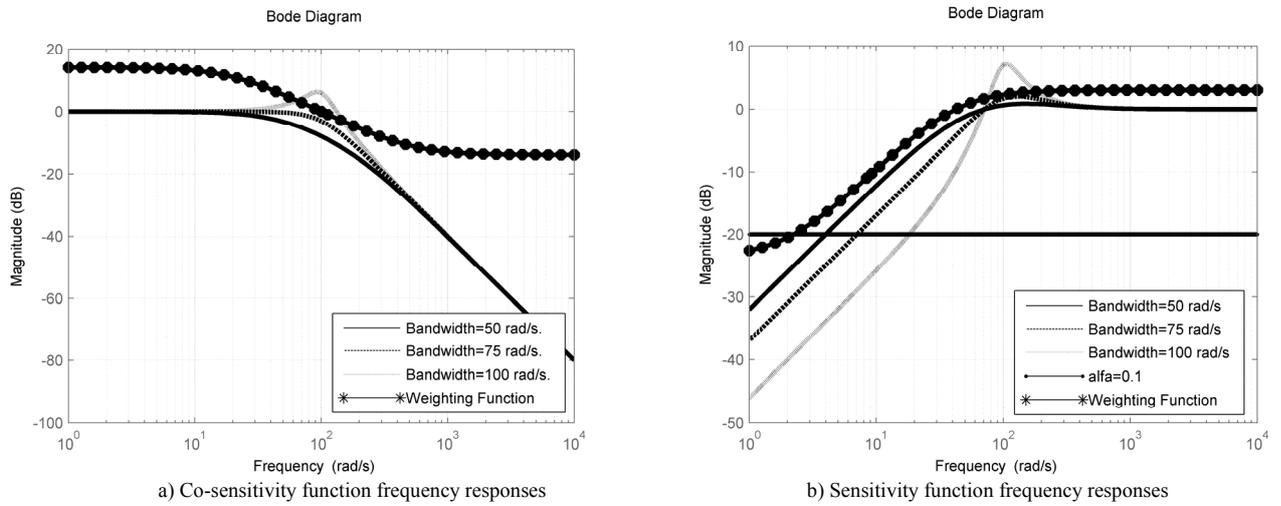

a) Co-sensitivity function frequency responses  b) Sensitivity function frequency responses

Fig. 5 Frequency responses of the inner-loop sensitivity and co-sensitivity transfer functions when a 2$^{nd}$ order DOB is used .

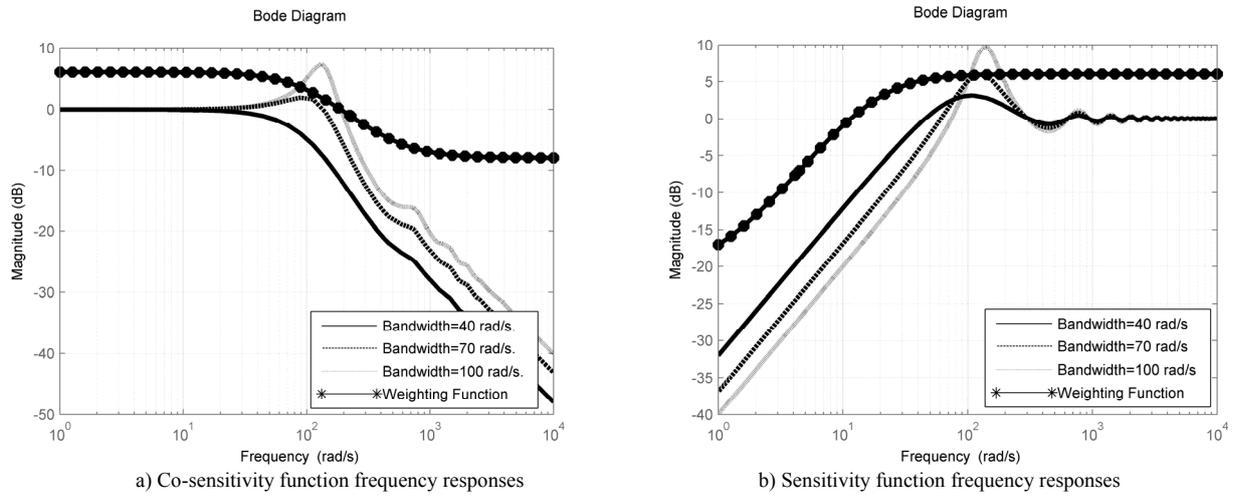

a) Co-sensitivity function frequency responses

b) Sensitivity function frequency responses

Fig. 6. Frequency responses of the inner-loop sensitivity and co-sensitivity transfer functions when the order of DOB is one.

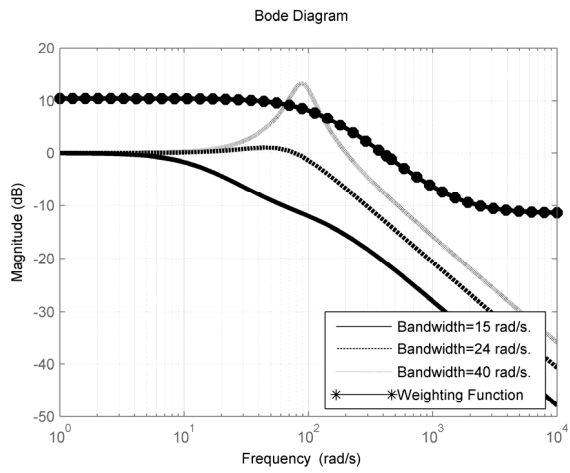
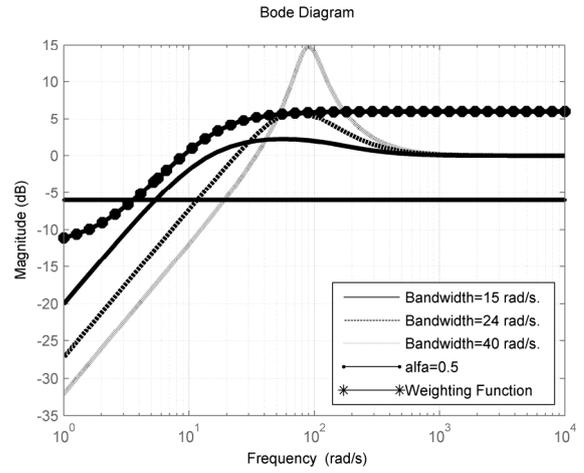

a) Co-sensitivity function frequency responses

b) Sensitivity function frequency responses

Fig. 7. Frequency responses of the inner-loop sensitivity and co-sensitivity transfer functions when the order of DOB is one.

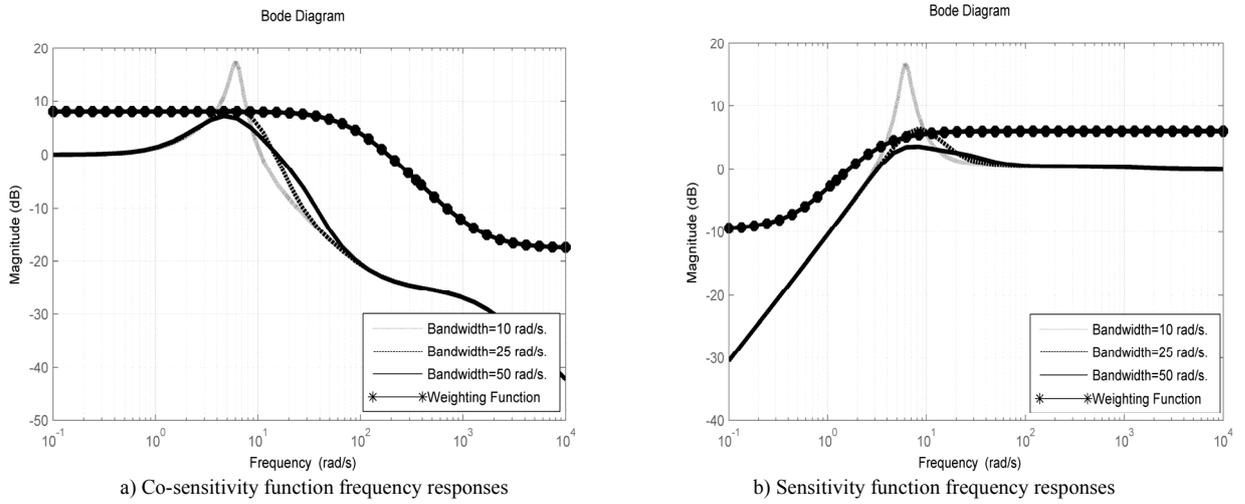

a) Co-sensitivity function frequency responses

b) Sensitivity function frequency responses

Fig. 8. Frequency responses of the outer-loop sensitivity and co-sensitivity transfer functions when the order of DOB is two

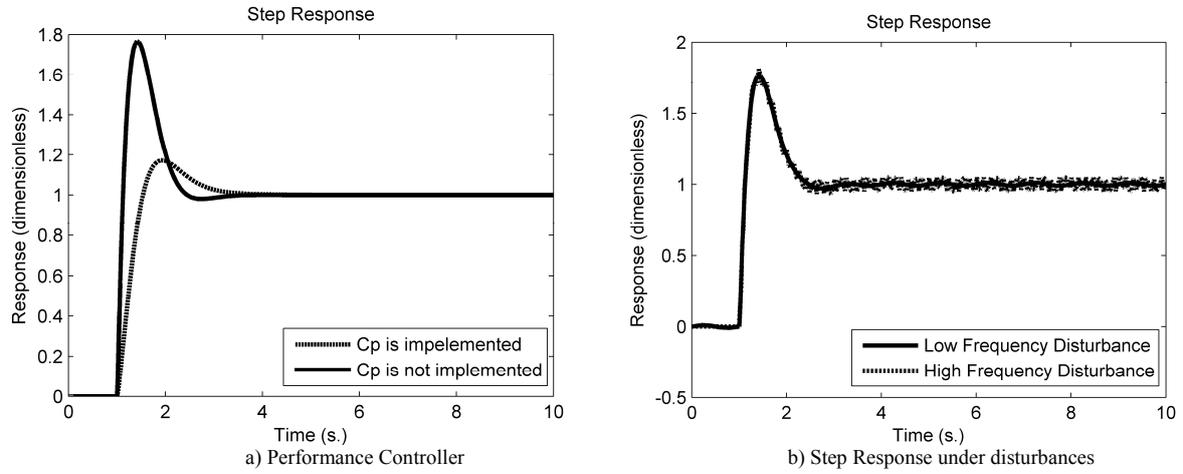

a) Performance Controller    b) Step Response under disturbances

Fig. 9. Step responses of the unstable plant when it is controlled by using the proposed robust controller

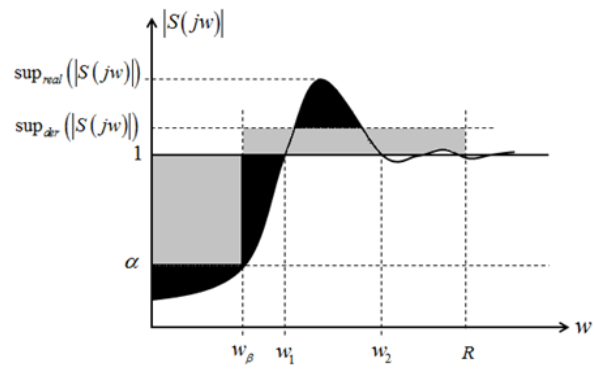

Fig. 10 A general frequency response of a sensitivity function